\documentclass[11pt]{llncs} 
\usepackage{graphicx}
\usepackage{caption}
\usepackage{amssymb} 
\usepackage{amsmath,amsfonts}
\usepackage{xspace}
\usepackage{color}
\usepackage{todonotes}
\usepackage{transparent}
\usepackage[vlined, linesnumbered, ruled]{algorithm2e}

\usepackage[a4paper, margin=1.2in]{geometry}

\newcommand{\block}[1]{\smallskip\noindent{\textbf{#1.}}~}

\newcommand{\oblot}{$\mathcal{OBLOT}$\xspace} 
\newcommand{\local}{$\mathcal{LOCAL}$\xspace} 
\newcommand{\congest}{$\mathcal{CONGEST}$\xspace} 
\newcommand{\Wait}{{\tt Wait}\xspace}
\newcommand{\Look}{{\tt Look}\xspace}
\newcommand{\Compute}{{\tt Compute}\xspace}
\newcommand{\Move}{{\tt Move}\xspace}
\newcommand{\LCM}{{\tt LCM}\xspace}

\newcommand{\A}{\mathcal{A}} 
\newcommand{\Ex}{\mathbb{E}}

\newcommand{\Aut}[1]{\mbox{Aut}({#1})}
\newcommand{\Canon}[1]{\mbox{Canon}({#1})}
\newcommand{\orbs}[1]{\mathcal{#1}} 
\newcommand{\oorbs}[1]{\bar{\orbs{#1}}} 
\newcommand{\NIL}{\mbox{NIL}}

\newcommand{\solve}{\mathsf{Solve}} 
\newcommand{\build}{\mathsf{Build}} 
\newcommand{\mtf}{\mathsf{MoveToFinal}} 
\newcommand{\move}{\mathsf{RobotMove}} 
\newcommand{\M}[1]{\mathcal{M}{(#1)}}

\newcommand{\fsync}{\textsc{FSync}\xspace}
\newcommand{\ssync}{\textsc{SSync}\xspace}

\newcommand{\async}{\textsc{Async}\xspace}

\definecolor{ser}{rgb}{0.95, 0.1, 0.1}
\definecolor{alf}{HTML}{0030f3}
\definecolor{ale}{HTML}{1d8348}
\definecolor{gab}{HTML}{dd13F8}

\newcommand{\problemnr}[4]{%
  \begin{problem}
  \label{#2}
  {#1}\vspace{1mm}
  \begin{list}{\ }{%
      \setlength{\topsep}{0ex}
      \setlength{\labelwidth}{4.5em}
      \setlength{\leftmargin}{5em}
      \setlength{\rightmargin}{0em}
      \setlength{\labelsep}{0.5em}
      \setlength{\itemsep}{0ex}
      }
      \item[\textit{Input:}\hfill] #3
      \item[\textit{Goal:}\hfill] #4 \medskip
  \end{list} 
  \end{problem}
}

\title{On the impact of unlimited computational power in \oblot: consequences for synchronous robots on graphs}

\author{Serafino Cicerone\inst{1} \and Alessia Di Fonso\inst{1} \and Gabriele Di Stefano\inst{1} \and \\ Alfredo Navarra\inst{2}}
\institute{Dipartimento di Ingegneria e Scienze dell'Informazione e Matematica, \\
        Universit\`a degli Studi dell'Aquila, Via Vetoio I-67100 
        L'Aquila, Italy. \\
\email{\{serafino.cicerone, alessia.difonso, gabriele.distefano\}@univaq.it}
\and
Dipartimento di Matematica e Informatica, 
        Universit\`a degli Studi di Perugia,\\ Via Vanvitelli I-06123 
        Perugia, Italy.
\email{alfredo.navarra@unipg.it}
}

\begin{document}

\maketitle

\begin{abstract}
The \oblot model has been extensively studied in theoretical swarm robotics.
It assumes weak capabilities for the involved mobile robots such as they are anonymous, disoriented, no memory of past events (oblivious), and silent. Their only means of (implicit) communication is transferred to their positioning, i.e., \emph{stigmergic} information. These limited capabilities make the design of distributed algorithms a challenging task.  Over the last two decades, numerous research papers have addressed the question of which tasks can be accomplished within this model. Nevertheless, as it usually happens in distributed computing, also in \oblot the computational power available to the robots is neglected as the main cost measures for the designed algorithms refer to the number of movements or the number of rounds required. 
In this paper, we prove that for synchronous robots moving on finite graphs, the unlimited computational power (other than finite time) has a significant impact. In fact, by exploiting it, we provide a definitive resolution algorithm that applies to a wide class of problems while guaranteeing the minimum number of moves and rounds.
\end{abstract}


\section{Introduction}
The \oblot model~\cite{FPS-macbook19} has been extensively studied in theoretical swarm robotics. Over the past two decades, many research papers have focused on determining whether a specific problem can be solved under the \oblot model and under which assumptions. Emphasis has been placed on feasibility issues rather than the computational power required by robots to accomplish a task.
In this model, a set of distributed and mobile robots has limited capabilities. 
They are anonymous, disoriented, no memory of past events (oblivious), and cannot communicate directly. The robots interact by observing the environment and the positions of all the other robots. Hence, robot interactions are based only on \emph{stigmergic} information.
The time is measured in \emph{rounds} for synchronous robots. In one round, any active robot performs a so-called \Look-\Compute-\Move (\LCM) cycle, whereas inactive robots are just idle. The subset of robots activated within one round either coincides with the whole set of robots (\fsync case), or is chosen by an ideal adversarial scheduler that provides some fairness, i.e., every robot is activated within finite time, infinitely often (\ssync case). 
In the \Look phase, a robot acquires a snapshot of the environment with information on the relative positioning of the other robots with respect to its own. During the \Compute phase, a robot makes computations -- without restrictions except for finite time -- to decide its next destination. 
Computations are based only on the snapshot acquired during the current \LCM cycle and on the designed algorithm, the same for all the robots. During the \Move phase, the robot moves toward the computed destination.

The robots need to coordinate to reach a shared goal. However, the limited robots' capabilities make the design of distributed algorithms a challenging task.
Many researchers have studied which tasks robots can perform within the \oblot model. They characterized various problems based on the initial configurations of robots, the type of time scheduler, and its variants.
The most relevant problems solved within the \oblot model concern \emph{Pattern Formation} issues~\cite{BAKS20,CDDN23,CDN19,CDN18c,GGSS23,SY99,YS10}, that is, robots must reach a final configuration (in terms of a particular disposal or of some property) within a finite time.
Standard tasks within this model concern, for instance, \emph{rendez-vous} or \emph{Gathering}~\cite{BKAS18,CDN20a,CFPS12,DDN14,DN17a,KKN10}, i.e. \emph{Point formation}, \emph{Mutual Visibility}~\cite{LunaFCPSV17,PoudelAS21,SharmaVT21} i.e., there must not be three collinear robots, and the \emph{Geodesic Mutual Visibility}~\cite{CDDN23-PMC} for robots moving on graphs in which between any couple of robots there must be a shortest path (a ``geodesic'') along which no other robots reside. 

The original \oblot model focused on robots moving on the Euclidean plane. In~\cite{KMP08}, the study of the Gathering problem has been initiated for robots constrained to move along the edges of ring network. 
One implicit assumption underlying the \oblot model, which the research community accepts, is that robots have unbounded computational power. However, this assumption does not imply that all problems are solvable in the \oblot model as the other constraints (i.e. anonymity, obliviousness\ldots) as well as the adopted time scheduler (i.e. synchronous, asynchronous) pose significant algorithmic challenges. 
In addition, in many distributed computing environments, which differ from standard algorithmic theory, cost metrics focus on the exchanged information rather than on the computational complexity. For example, in \emph{message-passing} models like \local and \congest~\cite{HS20,Linial92}, the emphasis is on the number (and the size) of the exchanged messages; in \oblot, on the number of robot movements. As a result, the robots' computational power is usually considered irrelevant.

\block{Our Results} In this paper, we prove that for robots moving on graphs, the assumption of an unbounded computational power has instead a significant impact. By leveraging this assumption, we provide a definitive resolution algorithm for synchronous robots moving on finite graphs that applies to a wide class $\mathcal{P}$ of problems while also optimizing the number of required moves and rounds.
The generality of our algorithm shows that it is important to limit the computational power of the robots to algorithms running in polynomial time when solving problems of robots moving on graphs.

In particular, we investigate feasibility issues in the \oblot model for robots moving on the vertices of finite graphs. We examine the \fsync scheduler.
As in \oblot, we assume that robots do not have any restrictions on the computational power during the \Compute phase. 
Through this approach, we develop an algorithmic framework that can be applied to a wide class of problems for determining if a given problem can be solved from a specific starting configuration. Furthermore, if the answer is yes, the algorithm also provides the minimum number of moves that robots must perform from each encountered configuration.
It introduces a ``configuration hypergraph'' to represent every potential transition between all possible configurations based on a specific input graph and a specific number of robots. This hypergraph enables us to derive a resolution algorithm through a step-by-step analysis computed by the robots.
A similar approach appears in~\cite{Potop-ButucaruSTU19}, where robot–environment interactions are modeled as a two-player reachability game on a graph~\cite{Games23}.
Our main result can be summarized by providing a general algorithm  that satisfies the following statement:
\begin{theorem}\label{th:preliminary}
    Let $\Pi \in \mathcal{P}$ be a problem in \oblot for synchronous robots moving on a finite graph. Starting from a configuration $C$, there exists an algorithm that allows the robots to reach a final configuration within the minimum number of rounds if and only if $C$ is solvable.
\end{theorem}
The full characterization of a given problem remains open, but our algorithm answers whether the problem can be solved for each input instance.

The importance of our findings relies in their theoretical implications. In the future, any attempts to solve problems under the \oblot model on graphs should consider limited computational power for the \Compute phase and polynomial in the size of the instance. Otherwise, our optimal decision algorithm provides a solution to many problems (including every variation of the pattern formation problem) from any given initial configuration. In addition, we make further considerations to generalize our result to different schedulers, for robots empowered with lights, and more.

\block{Outline}
In the next section, we revise in detail the \oblot model. In Section~\ref{sec:not}, we provide all the required notation for a correct formalization of our arguments. Section~\ref{sec:algo} presents the proposed algorithm.  Finally, Section~\ref{sec:concl} provides interesting concluding remarks. Furthermore, the Appendix contains useful discussions for the generalization to variants of the \oblot model.

\section{Robot model}\label{sec:model}
Robots are modeled according to $\mathcal{OBLOT}$ (see, e.g.,~\cite{FPS19} for a survey), one of the classical theoretical models for swarm robotics. In this model, robots are computational entities that can move in some environment (a finite graph in our case) and can be characterized according to a large spectrum of settings. Each setting is defined by specific choices among a range of possibilities, for a fundamental component - time synchronization - as well as other important elements, like memory, orientation, and mobility. We assume such settings at minimum as follows:
\begin{itemize}
\item \emph{Anonymous}: no unique identifiers;
\item \emph{Autonomous}: no centralized control;
\item \emph{Dimensionless}: no occupancy constraints, no volume, modeled as entities located on vertices of a graph;
\item \emph{Oblivious}: no memory of past events;
\item \emph{Homogeneous}: they all execute the same algorithm (no randomization features are allowed);
\item \emph{Silent}: no means of direct communication;
\item \emph{Disoriented}: no common reference system.
\end{itemize}

Each robot in the system has sensory capabilities allowing it to determine the location of other robots in the graph, relative to its location.
Each robot refers to a \emph{Local Reference System} (LRS) that might differ from robot to robot. 
%
Each robot follows an identical algorithm that is pre-programmed into the robot. 
The behavior of each robot can be described according to the sequence of four states: \Wait, \Look, \Compute, and \Move. Such states form a computational cycle of a robot.
\begin{enumerate}
\item \Wait. The robot is idle. A robot cannot stay indefinitely idle;
\item  \Look. The robot observes the environment by activating its sensors, which will return a snapshot of the positions of all other robots relative to its own LRS. Each robot is viewed as a point; 
\item  \Compute. The robot performs a local computation according to an algorithm $\A$ (we also say that the robot executes $\A$). The algorithm is the same for all robots, and the result of the \Compute phase is a destination point. 
Actually, for robots on graphs, the result of this phase is either the vertex where the robot currently resides 
or a vertex among those at one hop distance 
(i.e., at most one edge can be traversed within one cycle);
\item  \Move. If the destination is the current vertex where $r$ resides, $r$ performs a \emph{nil} movement (i.e., it does not move); otherwise, it moves to the adjacent vertex selected.
\end{enumerate}
When a robot is in \Wait, we say it is \emph{inactive}, otherwise it is \emph{active}. In the literature, the computational cycle is called the \Look-\Compute-\Move (LCM) cycle, as a robot is inactive during the \Wait phase. 

Since robots are oblivious, they have no memory of past events. This implies that the \Compute phase is based only on what is determined in their current cycle (in particular, from the snapshot acquired in the current \Look phase).  
Since robots refer to their own LRS, they cannot exploit a global reference system but their perception is based only on the relative positioning of the robots. 

Concerning the movements, in graphs, moves are always considered as instantaneous. This results in robots always being perceived on vertices and never on edges during the \Look phases. Hence, robots cannot be seen while moving.

Two or more robots can reside on the same vertex at the same time; this is called a \emph{multiplicity} (i.e., a vertex occupied by more than one robot). 
The robot's ability to perceive a multiplicity during the \Look phase is called the \emph{weak} multiplicity detection. Robots are considered to have the \emph{strong} multiplicity detection when they can detect the exact number of robots composing a multiplicity. Here we consider this latter case.

According to the literature, different  
characterizations are considered depending on whether robots are fully-synchronous, semi-synchronous, or asynchronous (cf.~\cite{FPS19,KKNPS24}). 
These synchronization models are 
defined as follows:
\begin{itemize}
\item \emph{Fully-Synchronous} (\fsync): Robots are always active and execute their \LCM-cycles in perfect synchrony. Time is divided into global rounds. In each round, robots take a snapshot, compute their move based on the snapshot, and execute it.
\item \emph{Semi-Synchronous} (\ssync): Robots are synchronized as in \fsync but
 not all robots are necessarily activated during a \LCM-cycle;
\item \emph{Asynchronous} (\async): Robots are activated independently, with each phase taking a finite but unpredictable time. Robots share no common notion of time.
\end{itemize}

In the \ssync and \async cases, the timing of the computational cycles is in the hands of the \emph{adversary}. This timing is assumed to be \emph{fair} meaning that every robot performs its \LCM-cycle within finite time and infinitely often. Without such an assumption the adversary may prevent some robots from ever moving.

The three synchronization schedulers induce the following hierarchy (see~\cite{DDFN18}): \fsync robots are more powerful (i.e., they can solve more tasks) than \ssync robots, which in turn are more powerful than \async robots. This follows by observing that the adversary can control more parameters in \async than in \ssync, and more in \ssync than in \fsync. That is, protocols designed for \async robots also work for \ssync and \fsync robots. Instead, any impossibility result for \fsync robots also holds for \ssync, and \async robots.  

Whatever the assumed scheduler is, the activations of the robots according to any algorithm $\A$ determine a sequence of specific time instants $t_0 < t_1 < t_2 < \ldots$ during which at least one robot is activated. In all types of schedulers, except for the \async case where robots do not share the notion of time, robots are synchronized. In the \fsync case, each robot is active at each time unit. In the \ssync, we assume that at least one robot is active at each time $t$.
If $C(t)$ denotes the configuration observed by some robots at time $t$ during their \Look phase, then an \emph{execution} of  $\A$ from an initial configuration $C$ is a sequence of configurations $\Ex: C(t_0),C(t_1),\ldots$, where $C(t_0)=C$ and $C(t_{i+1})$ is obtained from $C(t_i)$ by moving at least one robot (which is active at time $t_i$) according to the result of the \Compute phase as implemented by $\A$. 
%
\section{Preliminary concepts and notation}\label{sec:not}
A \emph{configuration} $C$ is modeled as $C=(G,\lambda)$, where $G$ is an undirected finite graph and $\lambda: V(G)\to \mathbb{N}$ provides the number of robots on each vertex of $G$ assumed that  $k=\sum_{v\in V(G)} \lambda(v)$ is bounded.

As said in the introduction, we devise an algorithmic framework that can decide whether a problem defined in the \oblot model can be solved starting from any given configuration. Our approach applies to a large class of problems. Let $\Pi$ be a generic problem defined in the \oblot model. Although $\Pi$ can be defined in different ways, for the sake of simplicity we first focus on \emph{formation} problems where a solution can be specified by a set $F$ of \emph{final configurations}.\footnote{Note that $F$ could, alternatively, be provided as input in the form of some property that the robots can verify on any configuration with $k$ robots located on $G$.} We call this class of problems $\mathcal{P}$. Notice that classical problems 
like \emph{Gathering},  \emph{Pattern Formation}, \emph{Mutual visibility} fit in $\mathcal{P}$.

Problem $\Pi\in \mathcal{P}$ depends on a finite graph $G$, on an integer $k\ge 1$ representing the number of robots located on $G$, and on a set $F$ of configurations $C=(G,\lambda)$, with $\sum_{v\in V(G)} \lambda(v)=k$. Note that the elements of $F$ are final for $\Pi$, that is, the displacement of robots in $C\in F$ is considered a solution for $\Pi$. In summary, $\Pi$ can be denoted by the triple $(G,k,F)$ and formalized as follows:

\problemnr{$\Pi=(G,k,F)\in \mathcal{P}$}{prob:gmv}{
Any  configuration $C=(G,\lambda)$ with $\sum_{v\in V(G)} \lambda(v)=k$.}%
{Design a distributed algorithm working under the \oblot model that, starting from $C$, brings all robots to a configuration $C'\in F$.
} 

In Section~\ref{sec:algo}, we provide a resolution algorithm called $\move$ that solves $\Pi=(G,k,F)$ as follows: if $\Pi$ can be solved starting from a configuration $C$, our algorithm guides all the robots to move so that (1) a configuration $C'=(G,\lambda')\in F$ is created, and (2) $C'$ is created with the minimum number of moves. If $C$ is a final configuration or $\Pi$ cannot be solved from $C$, then the algorithm does not make the robots moving (and each robot knows which of the two cases occurred).

In Section~\ref{sec:gen}, we discuss how to adapt the resolution algorithm for other kinds of problems and settings.
Due to the generality of $\Pi$, in the rest of this section, we introduce some concepts and notations necessary to formalize $\move$.

\block{Configurations and automorphisms} 
Two undirected graphs $G=(V,E)$ and $G'=(V',E')$ are \emph{isomorphic} if there is a bijection $\varphi$ from $V$ to $V'$ such that $\{u,v\} \in E$ if and only if $\{\varphi(u),\varphi(v)\} \in E'$. An \emph{automorphism} on a graph $G$ is an isomorphism from $G$ to itself. The set of automorphisms of a given graph $G$, under the composition operation, forms the automorphism group of  $G$, which is denoted as $\Aut{G}$. Two vertices $u$ and $v$ of $G$ are \emph{equivalent vertices} if there exists an automorphism $\varphi\in \Aut{G}$ such that $\varphi(u)=v$. The equivalence classes of the vertices of $G$ under the action of the automorphisms are called vertex \emph{orbits}. 

The concept of isomorphism can be easily extended to configurations. In our context,  $(G,\lambda)$ and $(G',\lambda')$ are isomorphic if there exists $\varphi\in \Aut{G}$ such that $\lambda(v)=\lambda'(\varphi(v))$ for each vertex $v$. As for graphs, $\Aut{C}$ is the automorphism group of $C=(G,\lambda)$. In $C$, two robots $r$ and $r'$ are \emph{equivalent robots} if they reside on vertices that are equivalent according to some automorphism of $\Aut{C}$. 
We denote by $\orbs{O}_C$ 
the set of the orbits of $C$ and by $\oorbs{O}_C$  the set of all the  orbits of $C$ with vertices occupied by robots.

\begin{figure}[t]
  \centering
  \resizebox{0.8\textwidth}{!}{%
       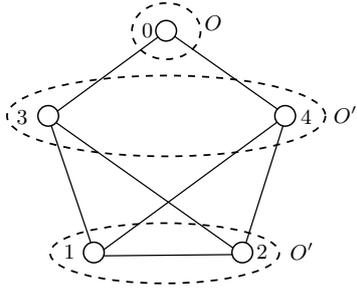    
       }
       \caption{An example of graph canonization, where dashed ovals enclose orbits. Note that $O < O' < O''$, according to the vertex labeling provided by the canonization.}
          \label{fig:canon}
\end{figure}

\block{Graph canonization} In the algorithms we provide below, all the robots must agree on some orbit to be used as a possible target where to gather. To this aim, we recall the \emph{graph canonization}, that is the process of finding a unique, standardized representation (called a canonical form) of a graph. The formal computational study of graph canonization began in the 1970s (e.g., see~\cite{BL83}). Formally, given a graph $G$, its canonical form is a graph $\Canon{G}$ such that for all graphs $G'$, $\Canon{G}=\Canon{G'}$ if and only if $G$ and $G'$ are isomorphic. For a graph $G$ with $n$ vertices, a canonical labeling algorithm will reassign integers from 0 to $n-1$ to vertices in a way that depends only on the structure of the graph (and not on the representation of the graph). As a consequence, the elements of $\orbs{O}_G$ can be totally ordered. In fact, given two orbits $O$ and $O'$, we say that $O < O'$ if, in the canonization of $G$, the minimum label of the vertices in $O$ is less than the minimum label of the vertices in $O'$. According to such a total ordering, all robots can agree on the \emph{smallest} orbit of $\orbs{O}_G$ (e.g., see Figure~\ref{fig:canon}). Quite recently, in~\cite{Babai19}, L{\'{a}}szl{\'{o}} Babai announced a quasi-polynomial-time algorithm for graph canonization, that is, one with running time $2^{O((\log n)^{c})}$ for some fixed $c>0$.

Notice that it is also possible to test whether the two configurations are isomorphic. It can be done, e.g., by transforming configurations into graphs so that two configurations are isomorphic if their corresponding graphs are isomorphic.
Let $C=(G,\lambda)$ be a configuration. We define its \emph{configuration graph} as the graph $\Gamma_C$ obtained by connecting $\lambda(v)+1$ new vertices to each vertex $v$ of $G$, as pendant vertices.
Formally:
$$ V(\Gamma_C)=V(G)\cup \bigcup_{v\in G} \{v_0, v_1,\ldots, v_{\lambda(v)}\}$$

$$ E(\Gamma_C)=E(G)\cup \bigcup_{v\in G} \{vv_0, vv_1,\ldots, vv_{\lambda(v)}\}$$

In so doing, the pendant vertices of a vertex $v$ in $\Gamma_C$ play the role of the robots occupying vertex $v$ in $C$. It follows that:

\begin{lemma}
Given two configurations $C=(G,\lambda)$ and $C'=(G',\lambda')$, then $C$ and $C'$ are isomorphic if and only if $\Canon{\Gamma_C}=\Canon{\Gamma_{C'}}$.
\end{lemma}

In what follows, for a given configuration $C$, we denote its canonization with $\Canon{C}$ instead of $\Canon{\Gamma_C}$.

\block{Moves}
Given a configuration $C=(G,\lambda)$, its orbits $\orbs{O}_C$
can be ordered since, given two orbits $O$ and $O'$, we can say that $O < O'$ if in the canonization of $C$, the minimum label of the vertices in $O$ is less than the minimum label of the vertices in $O'$. 
We say that two orbits $O$ and $O'$ (not necessarily distinct) are \emph{adjacent} if and only if there is an edge $xy$ in $G$ such that $x\in O$ and $y \in O'$.

Recalling that $\oorbs{O}_C=\{O_1,O_2,\ldots,O_t\}$ is the set of all the occupied orbits of $C$, a move in $C$ is a function $m_C:\oorbs{O}_C\to \orbs{O}_C\cup\{nil\}$ that specifies the adjacent orbit that the equivalent robots lying on $O$ have to reach,  for each orbit $O\in \oorbs{O}_C$. If $m_C(O)=nil$, for some $O\in \oorbs{O}_C$, the robots lying on $O$ do not move. Since the vertices in the destination orbit are equivalent, as well as the moving robots, the adversary decides the specific vertex that each robot reaches within the specified orbit. 
\begin{figure}[ht]
   \centering
   \def\svgwidth{0.9\columnwidth}
     {\fontsize{8}{10}\selectfont 
     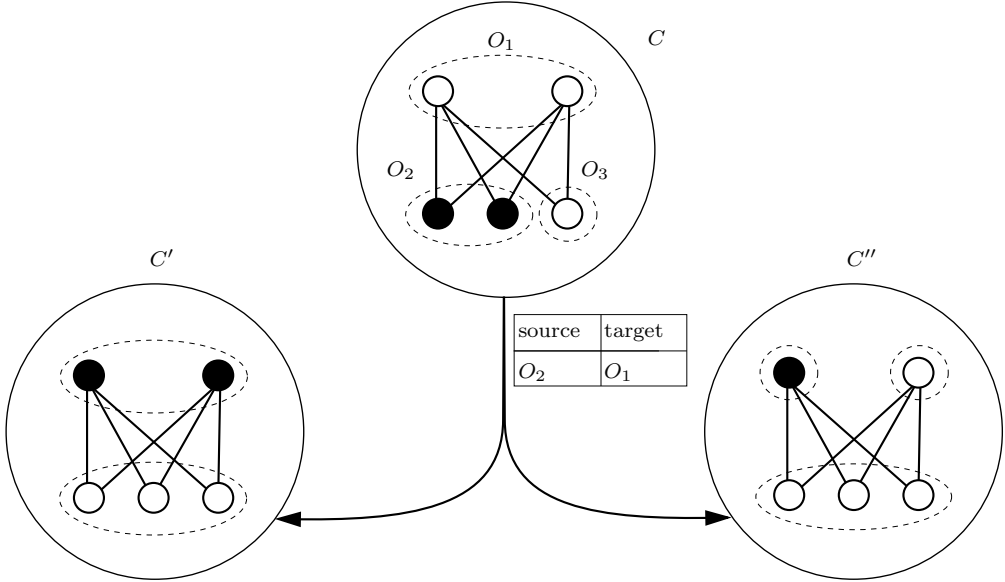
     }
    \caption{A configuration $C$ for a graph $G=K_{2,3}$ and $k=2$ robots. Dashed ovals enclose orbits. The figure shows two configurations reachable with a single move.}
\label{fig:move}
\end{figure}
Figure~\ref{fig:move} shows a configuration $C$ for the graph $K_{2,3}$ with $k=2$ robots both placed on orbit $O_2\in \oorbs{O}_C$. Then, the move that causes robots on $O_2$ to go to vertices in $O_1$ leads to one of the two different configurations $C'$ and $ C''$. The adversary decides the configuration that is really reached.

The $\NIL$ move is so that $\NIL(O) = nil$ for each orbit $O\in \oorbs{O}_C$ in any configuration $C$. A move $m_C$ can be seen as an ordered set of pairs $(O,m_C(O))$ for each orbit $O \in \oorbs{O}_C$. Hence, moves can be ordered since, given two moves $m$ and $m'$, we can say that $m<m'$ if the ordered sequence of pairs in $m$ is lexicographically less than the ordered sequence of pairs in $m'$. If $(O,m_C(O))$ is such that $m_C(O)$ is $nil$, we assume that the label associated with $nil$ is less than any other label used in the canonization of the configuration. 
\begin{figure}[ht]
   \centering
   \def\svgwidth{1\columnwidth}
   {\fontsize{7}{4}\selectfont 
     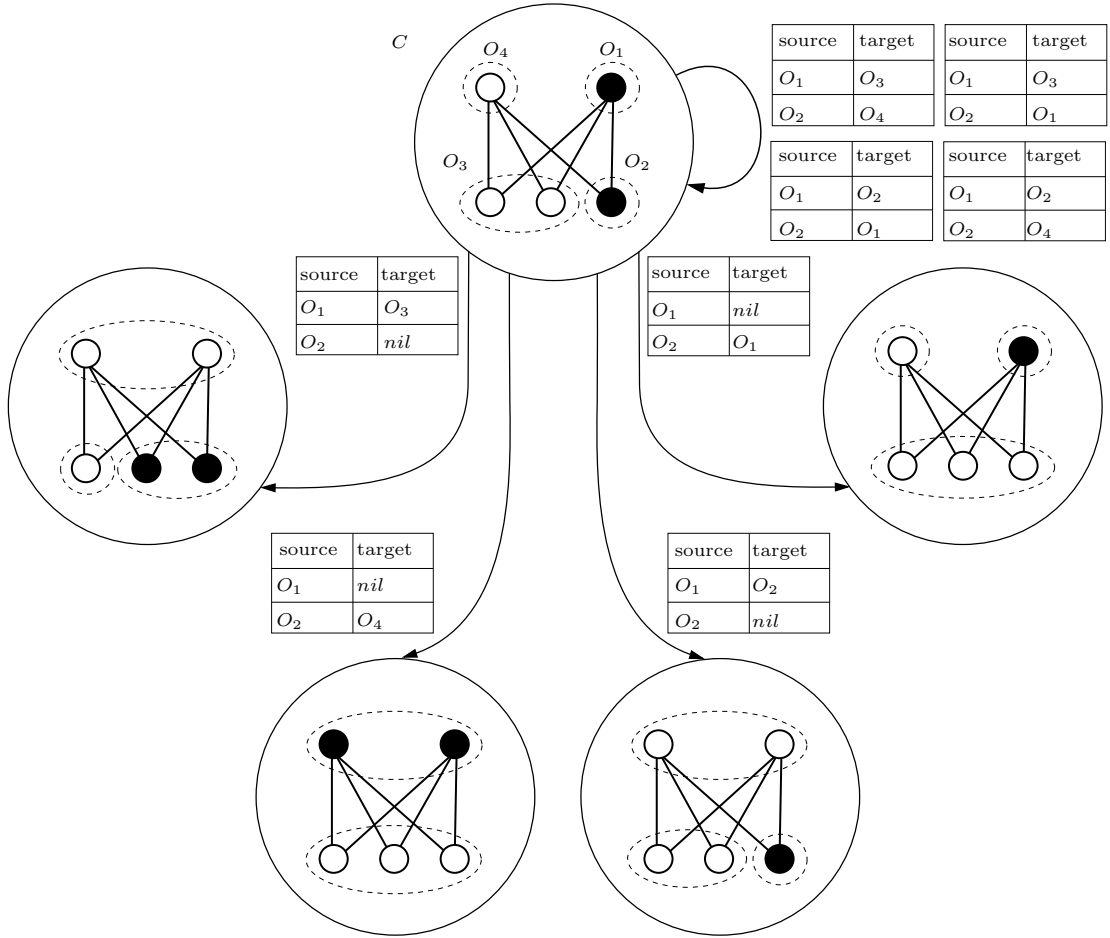
     }
    \caption{A configuration $C(G,k)$ for graph $G=K_{2,3}$ and $k=2$ robots and all the possible configurations reachable with a move. 
    }
\label{fig:conf}
\end{figure}
E.g., Figure~\ref{fig:conf} represents a configuration $C$ for the graph $K_{2,3}$ with $k=2$ robots placed on orbits $O_1$ and $O_2$. The hyperarc connecting $C$ to itself is associated with four moves having the same outcome.
The minimum move among them is the one that makes robots in $O_1$ move to $O_2$ and, concurrently, robots in $O_2$ move to $O_1$, whereas the minimum move among all the moves shown in the figure is the one where robots in $O_1$ stay in $O_1$ and, concurrently, robots in $O_2$ move to $O_1$.

\block{Configuration Hypergraph}
Let $G$ be a graph and $k$ be the number of robots displaced on the vertices of $G$. Starting from $G$ and $k$, it is possible to define a directed hypergraph  $H_{G,k}=(V,A)$ with vertex set $V(H_{G,k})$ containing all the configurations of $k$ robots on $G$, that is $V(H_{G,k})=\{C~|~C=(G,\lambda_C), \sum_{v\in V(G)} \lambda_C(v)=k \}$,  and with $(C,\Delta)\in A(H_{G,k})$ (the hyperarc set of $H_{G,k}$) whenever $C\in V(H_{G,k})$ and $\Delta\subseteq V(H_{G,k})$ is the set of all the configurations (up to isomorphisms) reachable from $C$ using a single move. We call $H_{G,k}$ the \emph{configuration hypergraph} of $G$ with $k$ robots.

\begin{figure}[ht]
   \centering
   \def\svgwidth{0.7\columnwidth}
     {\fontsize{10}{10}\selectfont 
\begingroup%
  \makeatletter%
  \providecommand\color[2][]{%
    \errmessage{(Inkscape) Color is used for the text in Inkscape, but the package 'color.sty' is not loaded}%
    \renewcommand\color[2][]{}%
  }%
  \providecommand\transparent[1]{%
    \errmessage{(Inkscape) Transparency is used (non-zero) for the text in Inkscape, but the package 'transparent.sty' is not loaded}%
    \renewcommand\transparent[1]{}%
  }%
  \providecommand\rotatebox[2]{#2}%
  \newcommand*\fsize{\dimexpr\f@size pt\relax}%
  \newcommand*\lineheight[1]{\fontsize{\fsize}{#1\fsize}\selectfont}%
  \ifx\svgwidth\undefined%
    \setlength{\unitlength}{722.68780085bp}%
    \ifx\svgscale\undefined%
      \relax%
    \else%
      \setlength{\unitlength}{\unitlength * \real{\svgscale}}%
    \fi%
  \else%
    \setlength{\unitlength}{\svgwidth}%
  \fi%
  \global\let\svgwidth\undefined%
  \global\let\svgscale\undefined%
  \makeatother%
  \begin{picture}(1,0.9813196)%
    \lineheight{1}%
    \setlength\tabcolsep{0pt}%
    \put(0,0){\includegraphics[width=\unitlength,page=1]{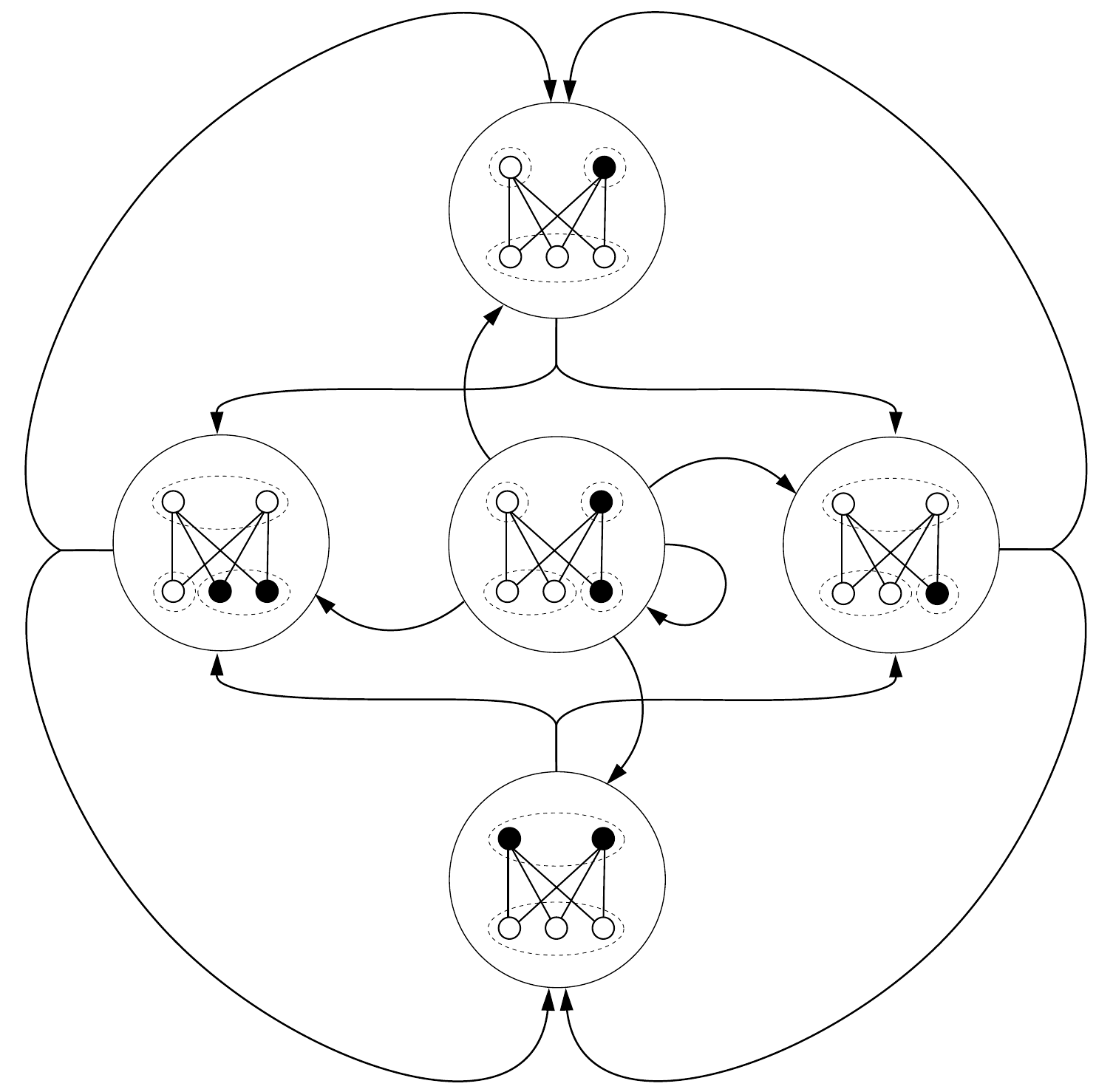}}%
    \put(0.53092614,0.89441303){\color[rgb]{0,0,0}\makebox(0,0)[lt]{\lineheight{1.25}\smash{\begin{tabular}[t]{l}$C_1$\end{tabular}}}}%
    \put(0.12307367,0.59656657){\color[rgb]{0,0,0}\makebox(0,0)[lt]{\lineheight{1.25}\smash{\begin{tabular}[t]{l}$C_2$\end{tabular}}}}%
    \put(0.48215015,0.59656657){\color[rgb]{0,0,0}\makebox(0,0)[lt]{\lineheight{1.25}\smash{\begin{tabular}[t]{l}$C_3$\end{tabular}}}}%
    \put(0.83499642,0.59656657){\color[rgb]{0,0,0}\makebox(0,0)[lt]{\lineheight{1.25}\smash{\begin{tabular}[t]{l}$C_4$\end{tabular}}}}%
    \put(0.43648712,0.29560659){\color[rgb]{0,0,0}\makebox(0,0)[lt]{\lineheight{1.25}\smash{\begin{tabular}[t]{l}$C_5$\end{tabular}}}}%
  \end{picture}%
\endgroup%

     }
    \caption{The configuration hypergraph of graph $G=K_{2,3}$ with $k=2$ robots. }
\label{fig:hyper}
\end{figure}
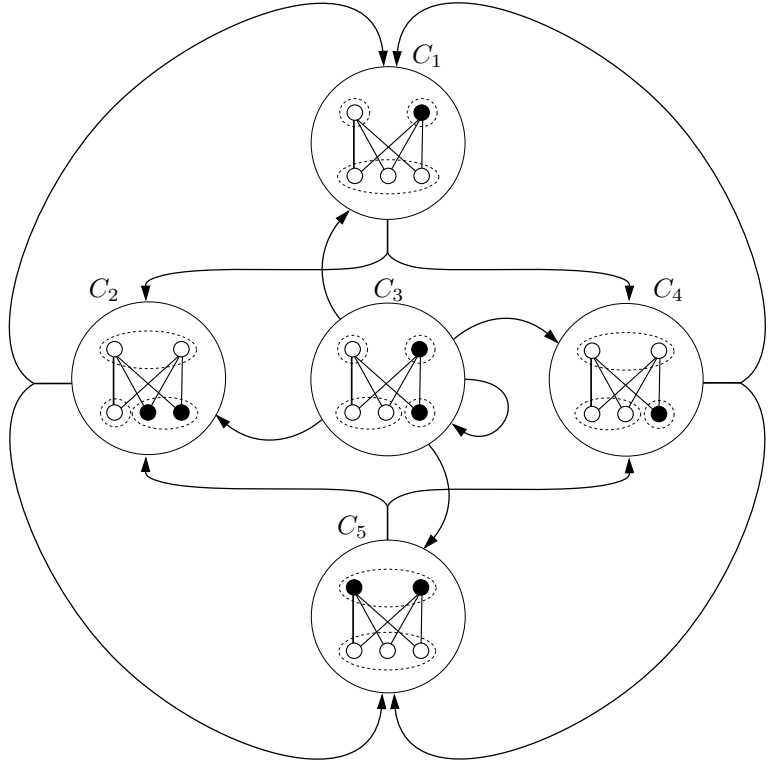

Let $(C,\Delta)$ be an hyperarc of $H_{G,k}$. Since each element of $\Delta$ can be reached from $C$ by applying different moves, we collect these moves in the set $\M{(C,\Delta)}$. Note that, given $C$ and $\Delta$, the set of moves in $\M{(C,\Delta)}$ is uniquely determined. The $\NIL$ move is never part of such a set. In Figure~\ref{fig:hyper}, the configuration hypergraph $H_{G,k}$ of the complete bipartite graph $G=K_{2,3}$ with $k=2$ robots is represented. In this example, 
$V(H_{G,k})$ is the set of the configurations $C_1,C_2,C_3,C_4$ and $C_5$, whereas $A(H_{G,k})$ is the set of the hyperarcs $(C_1,\{C_2,C_4\})$, $(C_2,\{C_1,C_5\})$, $(C_3,\{C_1\})$, $(C_3,\{C_2\})$, $(C_3,\{C_3\})$, $(C_3,\{C_4\})$, $(C_3,\{C_5\})$, $(C_4,\{C_1,C_5\})$, and $(C_5,\{C_2,C_4\})$. The moves associated with each hyperarc are not represented, but they should be clear since in each configuration but $C_3$ all the robots belonging to the same orbit can reach only one other orbit. As for configuration $C_3$, the hyperarcs and the associated moves are exactly those represented in Figure~\ref{fig:conf}.


\section{The algorithm}\label{sec:algo}
So far we are assuming robots in \fsync solving a problem $\Pi=(G,k,F)$ defined on a graph $G$ with $k$ robots and concerning a set $F$ of final configurations. A configuration $C$ is \emph{solvable} if it belongs to $F$ or there is a finite sequence of moves that from $C$ leads to a configuration in $F$.

Starting from a given configuration $C$, $\move$ (see Algorithm~\ref{algo:robot}) specifies the move that each robot in $C$ must perform to reach a final configuration in $F$ in an optimal number of moves, i.e., within the minimum number of rounds. Each robot runs Algorithm $\move$. It finds a move $m:\oorbs{O}_C\to \orbs{O}_C$ that associates a destination orbit $m(O)$ to each occupied orbit $O\in \oorbs{O}_C$. Then, each robot that recognizes itself as belonging to $O$ moves to a vertex in $m(O)$. Notice that the adversary selects the destination vertex in $m(O)$ when $|m(O)|\ge 2$.

\begin{algorithm}[t]
\small
\caption{ $\move(C=(G,\lambda), F)$} \label{algo:robot}
$k = \sum_{v\in G}\lambda(v)$\;
$H_{G,k}, \mathcal{M} = \build(G,k)$\;\label{l:build}
$S=\solve(H_{G,k},F)$\;\label{l:solve}
\If{$C\in S$}{ 
    $(d,m) = \mtf(C,H_{G,k},\mathcal{M}, F,S)$\;\label{l:mtf}
    perform move $m$\;\label{l:move}
       }
\Else{perform the \emph{NIL} move}\label{l:nil}
\end{algorithm}

Algorithm $\move$ receives, as parameters, a configuration $C=(G,\lambda)$ given in the canonical form and a set $F$ of final canonical configurations. It evolves as follows (see Algorithm~\ref{algo:robot}). At line~\ref{l:build}, using $\build$ (see Algorithm~\ref{algo:hypergraph}) it obtains the configuration hypergraph $H_{G,k}$ in which each possible configuration is represented in the canonical form and the corresponding set $\mathcal{M}$ of all the moves is associated to each hyperarc. Then, the set $S$ of all the solvable configurations is computed at line~\ref{l:solve} using Algorithm~\ref{algo:solve}, $\solve$.
If the input configuration $C$ is not solvable, the robot stays in place (see line~\ref{l:nil}), otherwise it finds a move $m$ that leads the current configuration $C$ toward a final configuration within a minimum number $d$ of rounds using Algorithm~\ref{algo:mtf}, $\mtf$ (see line~\ref{l:mtf}). 
Finally, the robot performs the move specified by $m$ as described above (line~\ref{l:move}).

\begin{algorithm}[t]
\small
\caption{ $\build(G,k)$ {\bf Output:} the configuration hypergraph, and the set of moves of each hyperarc } \label{algo:hypergraph}
$V=\{\Canon{C}~|~C=(G,\lambda_C), \sum_{v\in V(G)} \lambda_C(v)=k \}$\;\label{l:V}
$A=\emptyset$\;
\ForEach{ $C\in V$}{\label{l:for_C}
    Let $\orbs{O}_C^*$ be the set of orbits in $C$ adjacent to an orbit in $\oorbs{O}_C$\;\label{l:orbit_s}
    Let $\Omega=\{(O',O'') ~|~ O'\in \oorbs{O}_C, O''\in \orbs{O}_C^*\cup\{nil\}
    \}$\; \label{l:orbit_e}
    \ForEach{ $m\subseteq \Omega$ s.t. $m$ is a move}{ \label{l:move_s}
    generate the set $\Delta$ of all the configurations reachable from $C$ by $m$\;
    \If{$(C,\Delta)\not \in A$}{\label{l:if}
    add hyperarc $(C,\Delta)$ to $A$\;
    $\M{(C,\Delta)}=\{m\}$\;
      }
    \Else{
    add $m$ to $\M{(C,\Delta)}$\; \label{l:move_e}
      }
     }
    }
\Return (V,A), $\mathcal{M}$\; 
\end{algorithm}

Let us now describe algorithms $\build$, $\solve$, and $\mtf$. Starting from a graph $G$ and a number $k$ of robots, algorithm $\build$ constructs the configuration hypergraph $H_{G,k}=(V,A)$ having as vertex set $V$ all the possible configurations of $k$ robots on $G$ in the canonical form (see Algorithm~\ref{algo:hypergraph}). This is done at line~\ref{l:V}. The rest of the algorithm builds the set of hyperarcs $A$ and associates the corresponding set of moves to each hyperarc.
A hyperarc is a pair in which the first element is a configuration $C$ and the second element is a set of configurations reachable from $C$ by at least one move. Then, at line~\ref{l:for_C}, the algorithm starts a loop for each configuration $C$ to build all the hyperarcs having $C$ as first element.
At lines~\ref{l:orbit_s}--\ref{l:orbit_e} algorithm~$\build$ constructs the set $\Omega$ of all the pairs of two adjacent orbits also containing the pairs $(O',nil)$, for all the orbits $O'\in \oorbs{O}_C$. Any move $m$, seen as a set of pairs representing a function, is a subset of $\Omega$. Then, the loop from line~\ref{l:move_s} to line~\ref{l:move_e} first generates the set $\Delta$ of all the configurations reachable from $C$ by using the move $m$ and then adds the hyperarc $(C,\Delta)$ to the set $A$ and the move $m$ to the set $\M{(C,\Delta)}$ (see lines~\ref{l:if}--\ref{l:move_e}).  Eventually, the algorithm returns the configuration hypergraph $(V,A)$ along with the function $\mathcal{M}$ that associates the set of moves $\M{a}$ to each hyperarc $a\in A$.
\begin{algorithm}[t]
\small
\newcommand{\lIfElse}[3]{\lIf{#1}{#2 \textbf{else}~#3}}

\caption{ $\solve(H_{G,k}, F)$ {\bf Output:} all the solvable configurations} \label{algo:solve}

$S=F$\; \label{l:S}
$S'=\emptyset$\;

\While{$S\not = S'$}{\label{l:l}
  $S'=S$\;
  \ForEach{ $(C,\Delta)\in A(H_{G,k})$ }{\label{l:fa_s}
    \If{$\Delta\subseteq S'$\label{l:if1}}{
       $S=S\cup\{C\}$ \label{l:fa_e}
    }
  }
} 
\Return $S$\;     
\end{algorithm}

Algorithm $\solve$ finds all the configurations that are solvable. Note that, if there is a hyperarc $(C,\Delta)$ in the input hypergraph $H_{G,k}$ and all the configurations in $\Delta$ are solvable or are in the set $F$ of all the final configurations, then $C$ itself is solvable since it is sufficient to apply a move in $\M{(C,\Delta)}$ to transform $C$ into a solvable configuration in $\Delta$. Algorithm~\ref{algo:solve} at line~\ref{l:S} inserts the final configurations of $F$ in a set $S$ of solvable configurations. During the loop at line~\ref{l:l}, it makes a copy  $S'$ of $S$ and, at lines~\ref{l:fa_s}--\ref{l:fa_e}, adds to $S$ the configurations $C$ belonging to hyperarcs $(C,\Delta)$ such that all the configurations in $\Delta$ are solvable, that is $\Delta\subseteq S'$. The loop at line~\ref{l:l} ends when no new configurations are added to $S$, that is $S=S'$. E.g., in Figure~\ref{fig:hyper}, configurations $C_1$ and $C_4$ are final for the gathering problem. They compose the set $S$ at the beginning of the algorithm. Then, at the first step of the loop at line~\ref{l:l}, configuration $C_3$ is added to $S$ since there are two hyperarcs, namely $(C_3,\{C_1\})$ and $(C_3,\{C_4\})$, that connect $C_3$ to final configurations. At the second step of the loop, no new configuration is added to $S$, and the algorithm ends. Indeed, configuration $C_5$ cannot be added to $S$ since the only hyperarc starting from $C_5$ is $(C_5,\{C_2,C_4\})$ and $C_2$ is not solvable. Configuration $C_2$ cannot be added to $S$ since the only hyperarc starting from $C_2$ is $(C_2,\{C_5,C_1\})$ and $C_5$ is not solvable.
Then the algorithm correctly returns $S=\{C_1,C_3,C_4\}$.

\begin{theorem}\label{th:solve}
Algorithm $\solve$ outputs the correct set $S$ of all the solvable configurations.
\end{theorem}

\begin{proof}
We prove the claim by induction on the number of moves required by a configuration to reach a final configuration. Note that, the set of solvable configurations $S$ built iteratively by algorithm $\solve$ can only grow, as elements (configurations) are never removed. Such a set is initialized at line~\ref{l:S} by all the final configurations, which are solvable and require no movements.
Assume by induction that at the $i$-th iteration, $i\ge 0$, all the solvable configurations requiring at most $i$ moves have been correctly included in $S$. We prove that at the successive $(i+1)$-th iteration of the algorithm, any solvable configuration requiring $i+1$ moves is inserted in $S$. 
Indeed, at line~\ref{l:fa_e}, a configuration $C$ is inserted in $S$ if it admits at least a move that leads to a configuration $C'$ that was already in $S$ in the previous iteration, namely $S'$. 
In particular, at line~\ref{l:if1}, the algorithm checks whether a hyperarc leads to configurations in $S'$ (for each branch of the hyperarc). 
By the inductive hypothesis, $C'$ requires at most $i$ moves to reach a final configuration, hence $C$ requires at most $i+1$ moves. Furthermore, since by inductive hypothesis, any solvable configuration requiring at most $i$ moves has been correctly included in $S$, then any solvable configuration requiring $i+1$ moves will be detected by the algorithm as it explores all the hyperarcs at the  $(i+1)$-th iteration as well as at any other iteration.
\qed
\end{proof}

\begin{algorithm}[t]
\small
\caption{ $\mtf(C,H,\mathcal{M},\mathcal{V},F,S)$ {\bf Output:} number of moves and move to reach a final configuration}\label{algo:mtf}
\If{$C\in F$}{ \label{l:inF}\Return $(0,\NIL)$
       }
$M = \emptyset$\;
$\mathcal{V}$ = $\mathcal{V}\cup \{C\}$\;
\ForEach{ $(C,\Delta)\in A(H)$ s.t. $\Delta\subseteq S$ and $\Delta\cap \mathcal{V}=\emptyset$ }{\label{l:arcs}
    $d\_max=-1$\;\label{l:maxd_s}
    \ForEach{ $C'\in \Delta$ }{\label{l:loop_d}
       $(d,m)= \mtf(C',H,\mathcal{V},\mathcal{M},F,S$)\;\label{l:maxd_rec}
       \If{$d > d\_max$}{
         $d\_max=d$\;\label{l:maxd_e}
       }
    }
  $M$= $M \cup \left(d\_max,\min \M{(C,\Delta)}\right)$\; \label{l:M} 
}
$(d^*,m^*)=\min M$\;\label{l:min}
\Return $(d^*+1,m^*)$ \label{l:r}
\end{algorithm}

Given a solvable configuration $C$ in a configuration hypergraph $H$, Algorithm $\mtf$ returns the minimum number $d$ of moves to reach a final configuration $E\in F$ and the move $m$ that must be performed by the robots toward the next solvable configuration in $S$ along the shortest path to reach $E$.
The set $\mathcal{V}$ contains all vertices already visited, hence the first time a robot invokes $\mtf$ we have $\mathcal{V}=\emptyset$.
At line~\ref{l:inF}, the algorithm tests if the current configuration is final: in this case, it returns $0$ as the distance to a final configuration and $\NIL$ as the move to perform.
If the configuration is not final, the algorithm looks for a move by considering all the hyperarcs $(C,\Delta)$ such that all the configurations in $\Delta$ are solvable and none of them has been already encountered, i.e. $\Delta\subseteq S$ and $\Delta\cap \mathcal{V}=\emptyset$  (see loop at line~\ref{l:arcs}). Given such a hyperarc $(C,\Delta)$, since the adversary can choose the configuration $C'$ in $\Delta$ at maximum distance from a final configuration, this distance is computed at lines~\ref{l:maxd_s}--\ref{l:maxd_e} by recursively calling algorithm $\mtf$ on the configurations $C'$ in $\Delta$. In particular, this distance is stored in the variable $d\_max$. 
At line~\ref{l:M}, $d\_max$ is stored in $M$ along with the minimum move among those associated with edge $(C,\Delta)$. 
Choosing the minimum ensures that all the robots agree on the move to be performed if from $C$ the algorithm decides to reach a configuration in $\Delta$. 
At the end of the loop started at line~\ref{l:arcs}, the set $M$ contains a pair $(d,m)$ for each hyperarc considered in the loop, where $d$ is the maximum distance to a final configuration if the move $m$ is performed. Among these moves, the best pair $(d^*,m^*)$ is chosen at line~\ref{l:min} and returned with the correct distance $d^*+1$ at line~\ref{l:r}. 
E.g., if algorithm $\mtf$ is called from the solvable configuration $C_3$ of the hypergraph $H$ shown in Figure~\ref{fig:hyper}, since $C_3$ is not final, the algorithm considers the hyperarcs $(C_3,\{C_1\})$ and $(C_3,\{C_4\})$. The self-loop is clearly excluded by condition $\Delta\cap \mathcal{V}=\emptyset$.
For the hyperarc $(C_3,\{C_1\})$, a recursive call to $\mtf(C_1,H,\{C_3\},\mathcal{M},\{C_1,C_4\},\{C_1,C_3,C_4\})$ is performed, and since $C_1$ is final the pair $(0,\NIL)$ is returned. Then, at line~\ref{l:maxd_e}, in $d\_max$ is stored $0$. Hence, at line~\ref{l:M}, the pair $(0,m_{C_1})$ is stored in $M$, where move $m_{C_1}$ is the move to reach configuration $C_1$, making the robots in $O_1$ stay in $O_1$ and those in $O_2$ move to $O_1$ (see Figure~\ref{fig:move}). 
After a similar call for the hyperarc $(C_3,\{C_4\})$,  the pair $(0, m_{C_4})$ is also stored in $M$, where $m_{C_4}$ is the move to reach the configuration $C_4$ making robots in $O_1$ move to $O_2$ and those in $O_2$ stay in $O_2$. Then, at line~\ref{l:min}, $M=\{(0,m_{C_1}), (0,m_{C_4})\}$ and the minimum value $(d^*,m^*)$ is $(0,m_{C_1})$. Correctly, the value $(1,m_{C_1})$ is returned.

\begin{theorem}\label{th:mtf}
Given a solvable configuration $C$, Algorithm $\mtf$ outputs the maximum number of moves that from $C$ an adversary can force to reach a final configuration, along with the first move to perform.
\end{theorem}

 \begin{proof}
The proof proceeds by induction on the minimal number of moves required from $C$ to reach a final configuration. 
At the base of the induction, if $C$ is a final configuration, at line~\ref{l:inF} of Algorithm $\mtf$ such a condition is detected and the algorithm correctly returns a distance 0 to a final configuration, paired with the {\NIL} move.
We assume the claim is True for any solvable configuration with $d>0$ number of moves. Then, we prove the claim for $d+1$. 

Let $C$ be a solvable configuration requiring $d+1$ moves to be solved. This means that due to the adversary, no algorithm can guarantee to reach a final configuration in less than $d+1$ moves. 
For each hyperarc $(C,\Delta)\in A(H)$, the configurations reachable from $C$ with a single move $m$ are those in the set $\Delta$ with $m\in \M{(C,\Delta)}$. Since $C$ is solvable, there must exist at least a move in $\M{(C,\Delta)}$, such that all the configurations in $\Delta$ are solvable, that is $\Delta\subseteq S$, and none of them has already been encountered, that is $\Delta\cap \mathcal{V}=\emptyset$. All the moves that from $C$ lead to a solvable configuration are considered in the loop at line~\ref{l:arcs}.
Among such moves, since $C$ requires $d+1$ moves, there must exist at least a move $m'$ leading to a solvable configuration $C'$ that requires $d$ moves to be solved, with $d$ being the maximum number of moves that an adversary can force to perform from $C'$ to reach a final configuration. 
By the inductive hypothesis, the recursive call to $\mtf$ at line~\ref{l:maxd_rec} correctly returns a pair $(d,m')$, eventually. Any other pair $(d',m')$ returned by the other possible calls performed at line~\ref{l:maxd_rec} within the same iteration for the same set $\Delta$ must be such that $d'\leq d$. Indeed, $d'>d$ would mean that the adversary may force more than $d$ moves after completing the move that from $C$ leads to $C'$, against the hypothesis that $C$ requires at most $d+1$ moves to reach a final configuration. 
 At line~\ref{l:M}, the Algorithm stores the pair $(d,\min \M{(C,\Delta)})$ in the set variable $M$, recall that the set of moves $\M{(C,\Delta)}$ associated with each hyperarc $(C,\Delta)$ are ordered.
For any other pair $(d'',m'')$ inserted in $M$, computed in different iterations of the loop at  line~\ref{l:arcs}, due to other possible sets $\Delta'$ such that $\Delta'\subseteq S$ and $\Delta'\cap \mathcal{V}=\emptyset$ with $(C,\Delta')\in A(H)$, it is $d''\ge d$. 
Since by hypothesis, $C$ requires $d+1$ moves to reach a final configuration, there cannot exist an algorithm that, within one move, leads $C$ to a configuration that guarantees in less than $d$ moves to reach a final configuration.

Finally, at line~\ref{l:min}, the algorithm selects the pair $(d^*,m^*)$ as the minimum one in $M$, i.e. $d^*=d$ whereas $m^*$ is a move that leads from $C$ to some solvable configuration $C'$ that requires $d$ moves. The algorithm then returns the pair $(d^*+1, m^*)=(d+1,m^*)$ at line~\ref{l:r}, hence the claim holds.
\qed
\end{proof}

We are now ready to provide the proof and a more formal statement of Theorem~\ref{th:preliminary}:

\begin{theorem}\label{theo:main}
    Let $\Pi=(G,k,F)\in \mathcal{P}$. Starting from a configuration $C$, by repeatedly applying Algorithm $\move$, the synchronous robots reach a configuration in $F$ within the minimum number of rounds if and only if $C$ is solvable.
\end{theorem}
\begin{proof}
$(\Rightarrow)$: If by repeatedly applying Algorithm $\move$, the robots reach a configuration in $F$ within a minimum number of rounds then $C$ is solvable.

$(\Leftarrow)$:
As the first operation, Algorithm $\move$ computes the configuration hypergraph using Procedure $\build(G,k)$, see line~\ref{l:build}. Successively, according to Theorem~\ref{th:solve}, the algorithm correctly computes the set $S$ of all solvable configurations at line~\ref{l:solve}, by calling Procedure $\solve$. 

If $C\in S$, then at line~\ref{l:mtf} the algorithm calls Procedure~$\mtf$ that according to Theorem~\ref{th:mtf} correctly outputs a pair $(d,m)$, with $d$ being the maximum number of moves that from $C$ an adversary can force to reach a final configuration, whereas $m$ is the first move to perform. Hence, move $m$ is performed at line~\ref{l:move}. Of course, if $C\in F$, then $d=0$ and $m=NIL$.
Instead, if $d>0$ then the theorem ensures that the move $m$ performed leads to a configuration $C'$ from where the number of moves to be applied to reach a final configuration will be at most $d-1$.

Hence, in at most $d$ executions of Algorithm $\move$, the robots reach a configuration in $F$. Furthermore, the number of rounds required is the minimum according to an adversarial scheduler.
Finally, when $C\not \in S$, the given configuration is not solvable and the robots do not move, as specified in line~\ref{l:nil}, leaving the configuration unchanged. 
\qed
\end{proof}
%

\section{Challenging Generalizations}\label{sec:gen}
In this section, we discuss possible variants to the settings described so far, hence generalizing the scope of applicability of the proposed algorithm. 

\block{Different Scheduler}
Theorem~\ref{theo:main} states that for each problem $\Pi=(G,k,F)$ in $\mathcal{P}$, and for each solvable configuration $C$ for $\Pi$ composed of synchronous robots, Algorithm $\move$ can compute an optimal solution for $C$. Here, we briefly discuss why the proposed approach cannot be easily generalized to different schedulers like {\ssync} and {\async}. 

\begin{figure}[ht]
   \centering
   \def\svgwidth{0.8\columnwidth}
     {\fontsize{7}{10}\selectfont 
     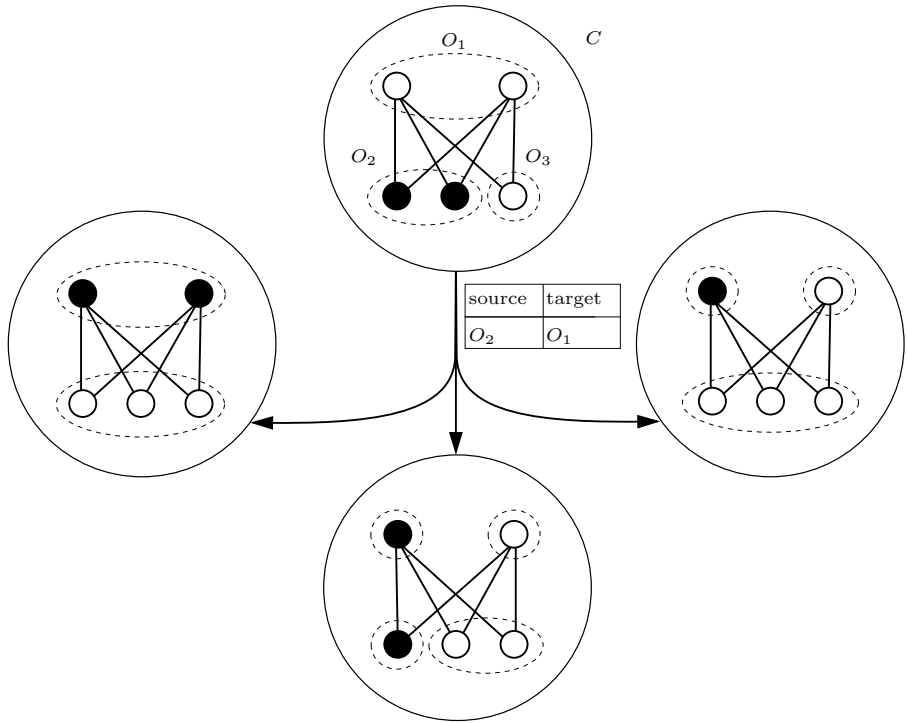
     }
    \caption{The same configuration $C$ shown in Figure~\ref{fig:move} along with the configurations reachable with a single move under the control of the \ssync scheduler.}
\label{fig:move2}
\end{figure}

\begin{figure}[ht]
   \centering
   \def\svgwidth{0.8\columnwidth}
\begingroup%
  \makeatletter%
  \providecommand\color[2][]{%
    \errmessage{(Inkscape) Color is used for the text in Inkscape, but the package 'color.sty' is not loaded}%
    \renewcommand\color[2][]{}%
  }%
  \providecommand\transparent[1]{%
    \errmessage{(Inkscape) Transparency is used (non-zero) for the text in Inkscape, but the package 'transparent.sty' is not loaded}%
    \renewcommand\transparent[1]{}%
  }%
  \providecommand\rotatebox[2]{#2}%
  \newcommand*\fsize{\dimexpr\f@size pt\relax}%
  \newcommand*\lineheight[1]{\fontsize{\fsize}{#1\fsize}\selectfont}%
  \ifx\svgwidth\undefined%
    \setlength{\unitlength}{722.68780085bp}%
    \ifx\svgscale\undefined%
      \relax%
    \else%
      \setlength{\unitlength}{\unitlength * \real{\svgscale}}%
    \fi%
  \else%
    \setlength{\unitlength}{\svgwidth}%
  \fi%
  \global\let\svgwidth\undefined%
  \global\let\svgscale\undefined%
  \makeatother%
  \begin{picture}(1,0.9813196)%
    \lineheight{1}%
    \setlength\tabcolsep{0pt}%
    \put(0,0){\includegraphics[width=\unitlength,page=1]{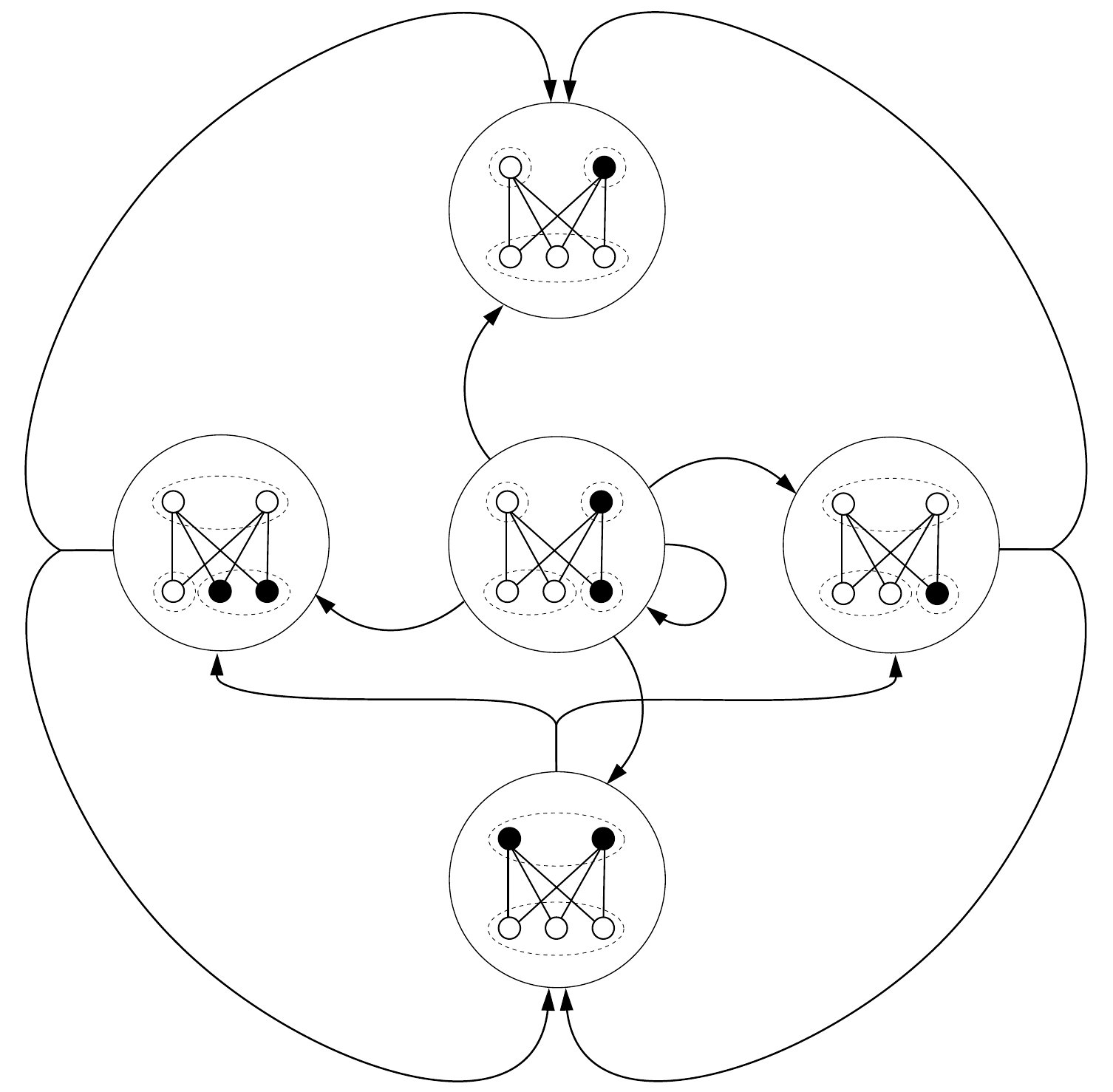}}%
    \put(0.53092614,0.89441303){\color[rgb]{0,0,0}\makebox(0,0)[lt]{\lineheight{1.25}\smash{\begin{tabular}[t]{l}$C_1$\end{tabular}}}}%
    \put(0.12307367,0.59656657){\color[rgb]{0,0,0}\makebox(0,0)[lt]{\lineheight{1.25}\smash{\begin{tabular}[t]{l}$C_2$\end{tabular}}}}%
    \put(0.44036478,0.59644068){\color[rgb]{0,0,0}\makebox(0,0)[lt]{\lineheight{1.25}\smash{\begin{tabular}[t]{l}$C_3$\end{tabular}}}}%
    \put(0.83499642,0.59656657){\color[rgb]{0,0,0}\makebox(0,0)[lt]{\lineheight{1.25}\smash{\begin{tabular}[t]{l}$C_4$\end{tabular}}}}%
    \put(0.43856271,0.29560659){\color[rgb]{0,0,0}\makebox(0,0)[lt]{\lineheight{1.25}\smash{\begin{tabular}[t]{l}$C_5$\end{tabular}}}}%
    \put(0,0){\includegraphics[width=\unitlength,page=2]{fig/allConfs_ssync_3.pdf}}%
  \end{picture}%
\endgroup%

    \caption{The configuration hypergraph for graph $G=K_{2,3}$ and $k=2$ robots under the control of the \ssync scheduler.}
\label{fig:hyperSS}
\end{figure}

Concerning the {\ssync} scheduler, each move generates more configurations than in the {\fsync} case. This is because, for every orbit $O$ occupied by robots, a move must be considered for any possible subset of robots in $O$. For example, referring to Figure~\ref{fig:move}, the same move under \ssync may result in one additional configuration, as illustrated in Figure~\ref{fig:move2}. In particular, the configuration shown at the bottom of the figure can occur if the adversary allows only one robot to move.
As a result, the entire configuration hypergraph shown in Figure~\ref{fig:hyper} for \fsync corresponds to that in Figure~\ref{fig:hyperSS} for \ssync. In practice, a major difference between {\fsync} and {\ssync} lies in the set of robots activated in each round by the adversary: in {\fsync}, all robots are activated, whereas in {\ssync}, a subset chosen by the adversary is activated.
It is worth noting that the data structures proposed in our approach can easily handle this larger number of generated configurations. Conversely, a specific characteristic of the {\ssync} scheduler — fairness — cannot be directly modeled in the solution we developed. As previously discussed, in the {\ssync} scheduler, the adversary determines which and how many robots are activated in each computational cycle, with fairness ensuring that each robot is activated infinitely often (without this assumption, the adversary could prevent some robots from ever moving).

Similarly, the proposed approach cannot be generalized to asynchronous robots, i.e. in {\async}. The main difficulty resides in managing the so-called {\em pending moves} (see, e.g.~\cite{CDN21a}). A move is called pending for a robot $r$ if $r$ has already acquired a snapshot during its \Look phase but has not yet performed the move while other robots may have already moved. Hence, the move that $r$ will eventually perform, is based on an outdated perception of the environment. Furthermore, when a robot takes its snapshot during its \Look phase, it cannot detect possible pending moves. It follows that under our approach, robots should guess the possible pending moves and all their combinations. Even though such a number is finite, and potentially a corresponding hypergraph can be constructed, a robot cannot detect the actual configuration on this hypergraph. As a result, it is impossible to determine a move that may lead to a final configuration because the starting point is undetectable.

\block{Different Capabilities}
The proposed algorithm can be applied to many variants of the \oblot model, as long as robots can detect, or at least cohesively guess, the current configuration on the configuration hypergraph.
For instance, in the $\mathcal{LUMI}$ variant (see, e.g.,~\cite{DDFN18,DFPSY16,FMP23}), robots are empowered with a visible light assuming a finite set of colors. This light can be used to communicate or to store information.
Hence, the construction of the hypergraph must also deal with the possible colors that each light associated with robots may assume. However, this remains computable and the size of the hypergraph remains finite.
Interestingly, as it has been proven in~\cite{DFPSY16}, any task solvable in {\ssync} with a constant number of lights is also solvable in {\async} with a constant number of lights.
In fact, lights can be used in {\async} to encode the status of a robot, hence the current configuration can be detected on the corresponding hypergraph.

\block{Different types of problems}
The proposed approach is applicable to a wider class of problems than the class $\mathcal{P}$. For instance, consider the {\em Patrolling} problem (see~\cite{CGK19a} and references therein), where a team of robots needs to visit all the vertices of a graph periodically. Once a robot has computed the configuration hypergraph as described above, it can analyze it to detect whether, from the current configuration, it is possible to enter a loop (of minimal length) that guarantees the visit of all the vertices of the underlying graph. For instance, looking at the configuration hypergraph of Figure~\ref{fig:hyper}, it can be deduced that two robots are insufficient to patrol $K_{2,3}$ since no sequence of moves can guarantee the visiting of all the vertices belonging to $K_{2,3}$ due to the symmetries of the graph.

\block{Infinite graphs}
The proposed algorithm explores all the possible moves that robots can make on the input graph. As a consequence, it cannot cope with infinite graphs as it is. However, every resolution algorithm must necessarily move robots within a finite portion of the input graph. Hence, the algorithm can be adapted to handle any problem where it is possible to limit a-priori the maximum distance traveled by the robots. For instance, the algorithms in~\cite{CDDN23,DN17} work on infinite grids but restrict the robot's movements within a predetermined portion of the grid. Furthermore, it is possible to find a resolution algorithm for a problem solvable on an infinite graph even if the maximum distance $d$ traveled by the robots is unknown. In this case, one can construct one configuration hypergraph for each increasing value of $d$ until finding the first hypergraph for which a solving algorithm is available. However, if the problem is not solvable, the computation would last forever as $d$ increases to infinity.


\section{Concluding remarks}\label{sec:concl}
We provided a definitive resolution algorithm for synchronous robots moving on graphs under the \oblot model. We have shown that the lack of constraints on the robot's computational capabilities allows the resolution of a wide range of problems while optimizing the number of required moves and rounds.

Note that, limiting the computational capabilities of robots may preclude the resolution of problems like the geodesic mutual visibility unless $P=NP$. Indeed, in~\cite{D22} it has been proven that computing the maximum number of robots that can be placed in mutual visibility on a graph is $NP$-hard, in general, even though there exist exact formulae for special graph classes~\cite{CiceroneDK23,D22}. 
Hence, limiting the robots' capabilities may lead to designing approximated solutions.

Clearly, the proposed algorithm raises a substantial warning by pointing out a weakness of the \oblot model that has not previously been explored. Interestingly, it also recommends importing standard approximation arguments in \oblot for certain problems.



\end{document}